\documentclass[prb, onecolumn,preprint,superscriptaddress]{revtex4-1}
\usepackage{graphicx}
\usepackage{latexsym}
\usepackage{amsmath}
\usepackage{amssymb}
\usepackage{epstopdf} 
\usepackage{enumerate}
\usepackage{setspace}   
\usepackage{dcolumn}
\usepackage{bm}
\usepackage{setspace}   
\usepackage{slashed}

\begin{document}
\title{Competition between superconductivity and nematic order: anisotropy of superconducting coherence length}

\author{Eun-Gook Moon }
\affiliation{Department of Physics, Harvard University, Cambridge MA
02138}
\affiliation{Department of Physics, University of California, Santa Barbara, CA 93106}

\author{Subir Sachdev }
\affiliation{Department of Physics, Harvard University, Cambridge MA
02138}

\date{\today}

\begin{abstract}
We study the interplay between  nematic order and superconductivity, motivated by a recent experiment on FeSe 
observing strongly distorted vortex shapes (Song {\em et al.}, Science {\bf 332}, 1410 (2011)).
We show that the nematic order strongly enhances the anisotropy in the superconducting coherence length, beyond
that expected from considerations of the Ginzburg-Landau theory. 
We obtain universal functions describing the coupling between the nematic order and superconductivity, and discuss
connections of our results to the experiments.
\end{abstract}

\maketitle

\section{Introduction}
The presence of multiple order parameters is one of the most interesting features of strongly correlated systems. 
For example, a number of order parameters appear in both the copper and iron based high temperature superconductors, including superconductivity and spin density wave  order (SDW).
Among them,  the electronic nematic order parameter, measuring spontaneous symmetry breaking of lattice rotation symmetry via
electron correlations, has been of particular interest in a series of remarkable experiments. \cite{hinkov08a,kohsaka07,taill10b,lawler,kim, ni08,nandi10}
In the cuprates, nematic order is signaled by the anisotropy of resistivity and bond ordered density\cite{hinkov08a,kohsaka07,taill10b,lawler,kim}. 
In the pnictides, orthorhombic structure distortion is ubiquitous, and 
is confirmed by various tools such as neutron scattering\cite{zhao}, STM\cite{chuang}, transport\cite{chu1,chu2,tanatar}, and ARPES\cite{shen}.
A natural question associated with nematic ordering is its interaction with other order parameters.
In superconducting materials, we are interested in whether the SC and nematic orders compete or attract.  
In the pnictides, and especially in $\rm{Ba_{1}[Fe_{2-x}Co_x]As_2}$ materials, it was shown that the nematic order parameter and SC  compete with each other, by observing the suppression of structural anisotropy 
along the onset of SC, the so-called ``back-bending'' \cite{ni08,nandi10,qcpshift} of the phase boundary.

Recently, Song {\em et al.} \cite{song} reported another interesting experiment 
in the  $\rm{FeSe}$ pnictides material showing striking interplay between nematic order and SC.
By observing scanning tunneling microscopy (STM) data of the zero bias resonance peak of the
electronic density of states,  
Song {\em et al.} found that shape of vortex cores was strongly distorted.
Even though the material has a orthorhombic structure, its structural anisotropy is much smaller than the 
anisotropy of vortices: the reported structural anisotropy is $0.5$ percent of the lattice constants, but the 
vortex shape anisotropy is order unity. 
Song {\em et al.} suggested other candidates, such as orbital ordering, to explain the large vortex anisotropy. 
Also, the experiments showed that the anisotropy is suppressed in strongly SC regions by observing essentially 
isotropic gap functions far from vortices.
Thus, it is clear that nematic order and SC compete each other.  
Near the vortex core, SC order parameter is suppressed, so nematic order is enhanced distorting vortex shapes significantly. 

This paper will present a general theory of competition between the SC and nematic orders, 
as concomitant instabilities of an underlying Fermi liquid. 
We will not address microscopic questions addressed in other literatures\cite{kivelson,cenke, seo, dagotto, fa, tesanovic,fa2,rafael}, 
for example, role of orbital physics in nematic ordering \cite{cwu}, but focus
on general results on the interplay between the two orders 
that follow from the symmetries of the order parameter and the presence of a Fermi surface.
We will show that the Fermi surface induces strong competition, which is especially important
in the vortex core, and which can greatly enhance the anisotropy in the vortex shape.
Non-analyticities associated with the Fermi surface will be shown to induce a large coupling
between the spatial gradients of the SC order and the nematic order parameter.
The influence of such a coupling on the vortex shape was studied recently \cite{dc} in the context of a Ginzburg-Landau
theory, and our results here offer a rationale for its enhancement.

Before introducing electrons and their Fermi surface, 
let us consider a simpler version of competing order parameter theory, so-called Ginzburg-Landau theory.
\begin{eqnarray}
\mathcal{F}& =& \frac{1}{2} (\partial \phi)^2 + \frac{r}{2} \phi^2 +\frac{u}{24} \phi^4 \nonumber \\
 &+& |\partial \psi|^2 + s |\psi|^2 +\frac{t}{4} (|\psi|^2)^2 \nonumber \\
 &+& \lambda |\psi|^2 \phi^2 + g_{GL} \phi \psi^{*}(-\partial_{x}^2 + \partial_{y}^2) \psi \nonumber \\
 &+& \cdots \label{GL}
\end{eqnarray}
Usual phenomenological constants for the GL theory ($r,u,d,y,\lambda,g_{GL}$) are introduced.\cite{dc} 
The first and second lines are for the nematic and superconducting order parameters.
The third line describes the coupling between the two order parameters. 
The coupling constant $\lambda$ characterizes how the two order parameters interact with each other : positive (negative) sign means  competition (attraction). 
One of the easiest ways to see the interplay is to observe the shift of the nematic order quantum critical point (QCP) shift due
to the presence of superconductivity:
\begin{eqnarray}
s_c(|\Delta|)=s_c(0) - \lambda |\Delta|^2.
\end{eqnarray}
Here, it is clear that the sign of $\lambda$ determines properties of interplay physics. Its positive(negative) value  indicates critical point shift toward(away from) the ordered phase, which shows competition(attraction) of the two order parameters.\cite{qcpshift,acl}
Let us consider another limit, where  the nematic ordering appears with the SC. 
The SC coherence length becomes anisotropic between the $x$ and $y$ directions 
\begin{eqnarray}
\xi_x= |s|^{-1/2}(1-g_{GL}\phi/2) \quad \xi_y= |s|^{-1/2}(1+g_{GL}\phi/2) \label{coherence}
\end{eqnarray}
So the $g_{GL}$ term describes the anisotropy$(\xi_x-\xi_y)/(\xi_x+\xi_y)$ of the SC coherence lengths and it is proportial to magnitude of the nematic order linearly. 
A small nematic order parameter implies small difference of coherence lengths in the GL theory.
Thus in this theory the structural orthorhombicity and nematic ordering cannot explain the 
strongly distorted vortex shape observed by Song {\em et al.}

We will discuss here the modifications of the GL theory described above due to the presence of electrons and a Fermi surface.
We will show below that the theory with electrons determines sign of the phenomenological constant $\lambda$, which turns out to describe the competition between the two order parameters. 
Also, we show that the coupling $g_{GL}$ corresponding to  anisotropic coherence length is significantly enhanced by 
the Fermi surface. Quite generally, such effects can be described by 
the one-particle irreducible vertex function in momentum space between two SC and one nematic order, 
\begin{eqnarray}
\Gamma_{\phi \psi^* \psi} = {\mathcal F}_{\phi \psi^* \psi} \left({\frac{p}{\Delta}} \right)  \frac{p_x^2 -p_y^2}{p^2},
\end{eqnarray}
, where $p_{x,y}, p, \Delta$ are for momentum components of SC order parameter, magnitude of the momentum, and SC gap magnitude.  
In the present terms, the GL theory clearly has 
\begin{eqnarray}
 {\mathcal F}^{GL}_{\phi \psi^* \psi} \sim p^2 \quad , \quad  \Gamma^{GL}_{\phi \psi^* \psi} \sim p_x^2 -p_y^2,
\end{eqnarray}
as follows directly from Eq.~(\ref{GL}).
Our main result of this paper is that the vertex function of the theory with electrons has a very different behavior
\begin{eqnarray}
{\mathcal F}_{\phi \psi^* \psi} \rightarrow const. \quad, \quad  \Gamma_{\phi \psi^* \psi} \sim \frac{p_x^2 -p_y^2}{p^2}.
\end{eqnarray}
in the $p/|\Delta|\rightarrow \infty$ limit, which is relevant the near the vortex core region due to SC suppression.
By comparing the two $\mathcal{F}_{\phi \psi^* \psi}$ functions at small momentum limit, it is manifest that anisotropy is much stronger once we consider electron degrees of freedom.

The non-analytic term above looks somewhat exotic, 
but it can be easily understood by extensions of the BCS theory in the presence of multiple order parameters. 
One way to understand this is to consider pairing susceptibility with infinitesimal SC pairing. 
Then, pairing susceptibility shows 
the well known ``BCS'' logarithmic behavior. \cite{mahan, tinkham}  
\begin{eqnarray}
\chi_{\rm pairing}=\langle\psi^{\dagger}(p) \psi(p) \rangle &\sim& \log(p^2),
\end{eqnarray}
$\psi ^{\dagger}(x)= c^{\dagger}_{\uparrow}(x) c^{\dagger}_{\downarrow}(x) $ is the pairing operator of electrons $c_\sigma$. 
In the presence of nematic order $\phi$, it is clear that there can be a correction to the pairing susceptibility of the form
\begin{eqnarray}
\chi_{\rm pairing}(\phi)=\langle \psi^{\dagger}(p) \psi(p) \rangle_{\phi} &\sim& \log(p^2 + \phi (p_x^2 -p_y^2))
\end{eqnarray}
The vertex function is readily obtained by differentiation with the nematic order parameter, 
\begin{eqnarray}
\Gamma_{\phi \psi^* \psi} \sim \frac{p_x^2 -p_y^2}{p^2},
\end{eqnarray}
which is the result above.
Below, we will calculate the full functional dependence of the vertex function with two parameters, the 
momentum and gap magnitudes. 

The main implication of the above considerations is that the linear dependence of the SC 
coherence length on the nematic order obtained in the GL theory is not correct.
Even small nematic order parameter can show strong enhancement due to its non-analytic influence.
Thus, we argue that such enhanced competition between SC and nematic order in the anisotropic coherence length channel contributes to experimentally observed vortex shape.

 The structure of this paper is as follows.
 In Sec. \ref{Model}, we set up our model Hamiltonian with fermions. 
We describe how the nematic and the superconducting order parameters couple to fermions and  introduce our strategy to study the competition effect. 
 For completeness,  we reproduce the quantum critical point shift of the nematic order parameter with fermions under superconductivity in Sec. \ref{Shift}. 
It is shown that the shift is not significant, in the sense that it is of the same order as predicted by the GL theory.
 Thus, for the QCP shift only, the fermions do not play an important role, and  the GL theory is enough to describe the competition physics. 
But we show that fermions do play an important role in the anisotropic coherence length of the SC order parameter in Sec. \ref{Aniso}. 
In Sec. \ref{Conclusion}, we discuss implication of our results.

\section{Model Hamiltonian and Strategy} \label{Model}
In two spatial dimensions, Ising nematic order 
on the square lattice breaks the four-fold rotational symmetry down to two-fold rotation. 
The Ising nematic order, could be realized microcopically by many routes, such as an orbital ordering which may be the case in the pnictides.
Instead of using orbital ordering however, we focus on a here simpler way to study the nematic order parameter- anisotropic hopping. 
 
Let us build a model Hamiltonian step by step. 
A non-interacting Hamiltonian on a square lattice is
\begin{eqnarray}
H_0= \sum_{i,j} -t_{ij} c_i^{\dagger} c_j ,
\end{eqnarray}
where $t_{ij}$ respects the fourfold rotational symmetry.
Because a nematic order parameter is coupled to an anisotropic hopping, the Hamiltonian with the nematic order parameter becomes
\begin{eqnarray}
H_1= \sum_{i,j} -t_{ij} c_i^{\dagger} c_j +\phi (c_i^{\dagger} c_{i \pm x}-c_i^{\dagger} c_{i \pm y}) + H_{nem}
\end{eqnarray}
The second term describes an anisotropic hopping, and by condensing the nematic order parameter, $\phi$,
the four-fold rotational symmetry(or $x,y$ exchange symmetry)  is broken down to two-fold. 
In momentum space, it is nothing but order parameter will distort the Fermi surface. 
The third term, $H_{nem}$, represents the nematic order parameter dynamics.

If we had used an orbital ordering model, the Fermi surface would contain 
orbital information, so that each momentum point is tied to the orbital direction.
This would modify matrix elements on the nematic order parameter in a momentum-dependent way,
but not crucially modify the Fermi surface physics we are interested in here.

Next, let us consider the superconducting order parameter.
In this paper, we mainly focus on the $s$-wave paired pnictides materials.
It is straightforward to generalize our results to other cases like $d$-wave pairing.
After a Fourier transform, the total Hamiltonian is  
\begin{eqnarray}
H_{tot}&=& \int_k( \epsilon(k) + \phi (\cos(k_x a) - \cos(k_y a) ) c(k) \nonumber \\
&+& ( \psi(p) c^{\dagger}(k+p) c^{\dagger}(-k)+ h.c.)+H_{nem} +H_{SC} ,
\end{eqnarray} 
where the spin index is suppressed. The $H_{SC}$ is the effective Hamiltonian 
for the superconducting order parameter. 
Note that, in the continuum limit, the nematic order parameter coupling becomes 
\begin{eqnarray}
\phi (\cos(k_x a) - \cos(k_y a) )  c^{\dagger}(k)c(k) &\rightarrow& \gamma(k) \, \phi \, c^{\dagger}(k)c(k) \nonumber \\
\gamma(k) &=&   \frac{1}{2} a^2 (k_x^2 -k_y^2) .
\end{eqnarray}

The total action is
\begin{eqnarray}
{\mathcal S}_{tot} &=& \int \Big[ \mathcal{L}_{\phi} + \mathcal{L}_{\psi} + c^{\dagger}_k (\partial_{\tau} +\epsilon(k)) c_k +\phi \gamma(k) c_k^{\dagger} c_k + (\psi_p c^{\dagger}_{k+p} c^{\dagger}_{-k} +  h.c. ) \Big]  
\end{eqnarray} 
The first two terms are for the order parameter dynamics, and one can understand these terms as usual $\varphi^4$ field theories with one and two components.
The remaining terms describe the fermion spectrum and its coupling to the two order parameters.
Here we absorb coupling constants of the Yukawa-type vertices into the order parameters. 
As expected, the two order parameters are coupled to each other by mediating electrons. 
Below, we study the total Lagrangian and how this theory describes the competition between the nematic order and the superconductivity. 

To see this, let us introduce some notations. Schematically, the effective action is expressed as follows. 
\begin{eqnarray}
Z&=& {\rm tr} (e^{- H/T}) =\int_{\phi,\psi} e^{-\Gamma_{eff}} \nonumber \\
\Gamma_{eff} &=& \int  \Gamma_{\psi^* \psi} \psi^{*} \psi + \Gamma_{\phi^* \phi} \phi^{*}  \phi+\Gamma_{\phi \psi^* \psi} \phi \psi^{*} \psi +\cdots. \label{Effective}
\end{eqnarray}
In the previous works\cite{acl,qcpshift}, the competition between SC and other order parameters was described by focusing on terms such as  the first two terms in Eqn. \ref{Effective}.
There, it was concluded that the SC and nematic order do not compete significantly, just as in GL theory, compared the competition between SC and spin density wave order (SDW).
In this paper, we focus on the third term($ \Gamma_{\phi \psi^* \psi} $) of the above equation, which basically describes anisotropy of coherence lengths. 

In this setup, one might argue that one can forget about the fermions in the low energy theory, expecting that the SC order parameter induces an energy gap near the Fermi surface. Then it seems  the Ginzburg-Landau theory of SC and nematic order, Eqn. \ref{GL}, would be enough to describe the system. However, it turns out that
such expectation is too naive because we are focusing on physics at non-zero momentum. So we need to keep the fermions to see the interplay physics between the order parameters, as shown below. 

\section{Nematic Quantum Critical Point Shift} \label{Shift}

In this section, we consider the nematic QCP shift under the SC.
The analysis is almost the same as in the previous work,\cite{qcpshift}, 
and the main difference is that we are consider whole Fermi surface, instead of hot-spots. 

Let us consider  the quantum critical point as that associated with the onset of nematic order, ${\phi}$, in a metal. This critical point will appear at a value $s_c^0$ of some tuning parameter $s$. We define $s$ so that $s < s_c^0$ is the nematic ordered phase with $\phi \neq 0$.(see the GL theory above).
The value of $s_c^0$ depends on material's microscopic detailes obviously. Then, let us consider  the onset of nematic order within
the superconductor characterized by a gap amplitude $\Delta$, and denote the critical
value of $s$ by $s_c(|\Delta|)$. The essence of the picture of competing orders is that the onset of superconductivity
should shrink the region of SDW order, and hence $s_c^0 > s_c(|\Delta|)$. 

Then, we evaluate the critical point shift($s_c^0 - s_c(|\Delta|) $) from coupling to the fermionic degrees of freedoms. It is determined by fermion susceptibility of the nematic order with and without SC. 
It is easy to obtain the susceptibility with SC, 
\begin{eqnarray}
\chi_{nem}  = \int_{k} \gamma(k)^2 \frac{|\Delta|^2}{(\epsilon(k)^2 +|\Delta|^2)^{3/2}}.  
\end{eqnarray} 
This integration looks proportional to magnitude of the SC order parameter, $|\Delta|$. However the integral
is non-zero as $\Delta \rightarrow 0$, because of the finite density of states at the Fermi level. 

Here we expand the anisotropy term around the Fermi surface, and take the lowest term, ($\gamma(k) \sim \frac{g}{2} \cos(2\theta_k)$) for simplicity.
Higher term corrections are subdominant, once we take the BCS limit($\Delta/E_F \ll 1$). 
Then we obtain 
\begin{eqnarray}
\chi_{nem} = \frac{g^2 \mathcal{N}(0)}{2} (1-\alpha_2|\Delta|^2)\quad , \quad s_c^0 - s_c(|\Delta|)=\frac{g^2 \mathcal{N}(0)}{2}\alpha_2|\Delta|^2
\end{eqnarray}
where $\mathcal{N}(0)$ is a density of states at the Fermi energy, and $\alpha_{2}$ is a cutoff-dependent positive constant. 
Therefore, the presence of SC suppresses the nematic order susceptibility, which indicates competition between the two order parameters. 
The $\alpha_2$ term corresponds to the $\lambda$ term in the GL theory in Eqn. \ref{GL}. 
This calculation shows that a non-analytic term does not appear in the nematic order QCP shift ,
and the competition effect is relatively weak compared to the linear QCP shift as in spin density wave(SDW).\cite{qcpshift} 
In other words, the competition basically follows the GL theory, so it is not necessary to keep fermions to understand the QCP shift of the nematic order parameter under SC. 

In the next section, we show that a non-trivial coupling appears in a finite momentum coupling corresponding to anisotropic SC coherence lengths by Fermi surface contributions, and the coupling is much stronger than usual GL couplings in the low momentum limit.

\section{Enhanced anisotropy of SC coherence lengths} \label{Aniso}

\begin{figure}
\includegraphics[width=2.2in]{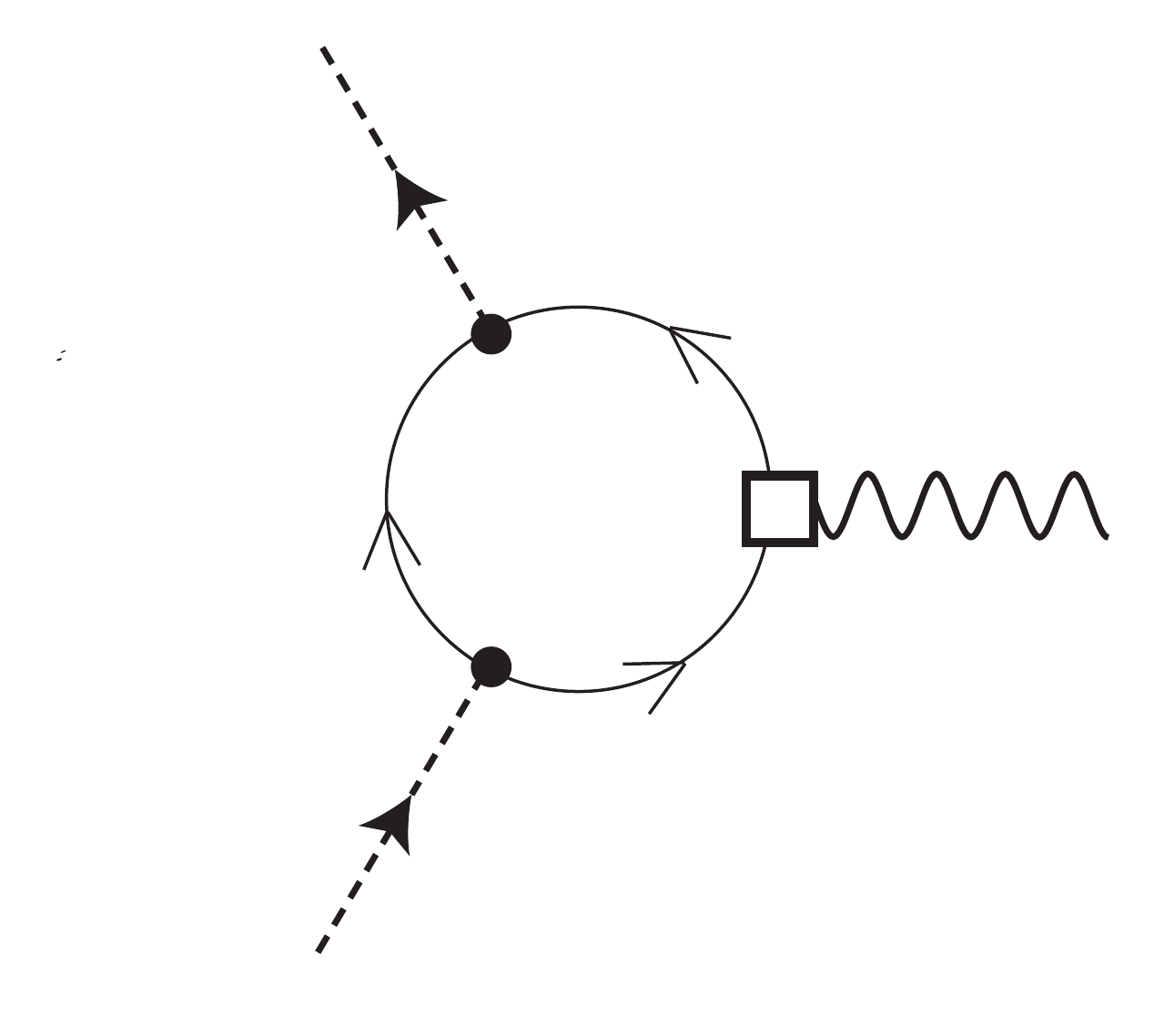}
\caption{Feynman diagram for the anisotropic coherence length vertex. 
The wavy line is for the nematic order parameter and the dashed lines are for the SC. The plain lines of the loop is for electrons's propagators with pairing. 
Note that directions of the fermion propagator is determined by the normal component of propagator.  } \label{loop}
\end{figure}

To incorporate the fermion contribution to anisotropic SC coherence lengths, we need to evaluate one Feynman diagram for the vertex term($ \Gamma_{\phi \psi^* \psi} $) as in Fig. \ref{loop}.
The electron propagator contains anomalous terms from SC pairing. 
The uniform SC order parameter is represented by $|\Delta|$, and we consider the non-zero momentum component of SC modes.  
Also, in this paper, we only consider uniform nematic order parameter.  
In this case, it is easy to show that the vertex evaluation is equal to derivatives of the pairing susceptibility. 

Without loss of generality, we can assume that SC order parameter can be adjusted as a real number by choosing a proper gauge.  
In the SC phase, the pairing susceptibility depends on its direction. 
\begin{eqnarray}
&&\chi_{xx}(2p) = \int_k \frac{1}{E_{k+p}+E_{k-p}}(1+\frac{\varepsilon_{k+p}\varepsilon_{k-p}-\Delta_{0}^2}{E_{k+p}E_{k-p}}) \nonumber \\
&&\chi_{yy}(2p) = \int_k \frac{1}{E_{k+p}+E_{k-p}}(1+\frac{\varepsilon_{k+p}\varepsilon_{k-p}+\Delta_{0}^2}{E_{k+p}E_{k-p}})
\end{eqnarray}
We define  relations $\chi_{xx} = \chi_0 -\delta \chi$ and  $ \chi_{yy} = \chi_0 +\delta \chi $ so that
\begin{eqnarray}
S_{eff}^{pairing} = \int \chi_0 |\psi|^2 + \delta \chi (\psi \psi + \psi^* \psi^*)+\cdots
\end{eqnarray}

If we assume that the band structure $\varepsilon(k)$ contains the nematic order parameter contribution, 
then it is easy to obtain the SC coherence anisotropy vertex by differentiating the pairing susceptibilies with nematic order parameter. 
Assuming $\varepsilon(k)$ is isotropic, the nematic order parameter just renormalizes a dispersion relation as follows.  
\begin{eqnarray}
\varepsilon(k) \rightarrow \varepsilon(k)+ \phi \gamma(k^2) \cos(2\theta_k) ,
\end{eqnarray} 
where $\gamma(k^2)$ is for an isotropic magnitude dependent part. Then,
\begin{eqnarray}
S_{eff}^{pairing} = \int \cdots+\Gamma_{\phi \psi^* \psi}   \phi |\psi|^2 +\tilde{\Gamma}_{\phi \psi^* \psi} \phi (\psi \psi + \psi^* \psi^*) +\cdots
\end{eqnarray}
Full expressions for the vertex functions are 
\begin{eqnarray}
\Gamma_{\phi \psi^* \psi} =\frac{\partial \chi_0}{\partial \phi}(p)|_{\phi=0}&=&2\int_k \frac{-1}{(E_{k+p}+E_{k})^2}(1+\frac{\varepsilon_{k+p}\varepsilon_{k}}{E_{k+p}E_{k}})( \frac{\varepsilon_{k}}{E_{k}} \frac{\partial \varepsilon_{k}}{\partial \phi} )+ \frac{\Delta_0^2}{E_{k+p}+E_{k}}(\frac{\partial \varepsilon_{k}}{\partial \phi}  \frac{\epsilon_{k+p}}{E_{k+p} E_{k}^3}) \nonumber \\
\tilde{\Gamma}_{\phi \psi^* \psi} =\frac{\partial \delta \chi}{\partial \phi}(p)|_{\phi=0}&=&2\int_k \frac{-1}{(E_{k+p}+E_{k})^2}\frac{\Delta_0^2}{E_{k+p}E_{k}}( \frac{\varepsilon_{k}}{E_{k}} \frac{\partial \varepsilon_{k}}{\partial \phi} )+ \frac{-\Delta_0^2}{E_{k+p}+E_{k}}(\frac{\partial \varepsilon_{k}}{\partial \phi}  \frac{\epsilon_{k}}{E_{k+p} E_{k}^3}).
\end{eqnarray}
Here, we only focus on the BCS limit($\Delta_0 /E_F \rightarrow 0$). 

So far, our setup is very general. 
But to extract more information, let us introduce three parameters ($C_1, C_2, C_3$) for the electronic density of states, effective mass, and the nematic coupling. 
With a general band structure, we can assume that momentum shifted dispersion relation is 
\begin{eqnarray}
&&\varepsilon_{k+p} = \varepsilon_k + v_F p \cos(\theta_{k}-\theta_{p}) +C_2 \frac{p^2}{E_F}, 
\end{eqnarray}
where $C_2$ is a non-universal number determining the effective mass. 
In general, the Fermi velocity has angular dependence with the four-fold rotational symmetry, and nematic order parameter breaks the symmetry down to the two-fold one.  
In this paper, we assume the Fermi velocity is $v_F$ is isotropic and set to unity. Its angular dependence only changes numerical factors of the final results, specifically the constants ($C_1,C_2,C_3$) of Eqn. \ref{ftns}.  

Also, we assume that the density of states at the Fermi surface is constant
\begin{eqnarray}
\mathcal{N}(E_F) = C_1 E_F,
\end{eqnarray}
where $C_1$ is a band structure related constant. 

Then, the above vertex functions are
\begin{eqnarray}
\Gamma_{\phi \psi^* \psi} (p)&=& 2 C_1 E_F \int \frac{d \theta}{2 \pi} \int d{\varepsilon}  \frac{-1}{(E_{k+p}+E_{k})^2}(1+\frac{\varepsilon_{k+p}\varepsilon_{k}}{E_{k+p}E_{k}})( \frac{\varepsilon_{k}}{E_{k}} \frac{\partial \varepsilon_{k}}{\partial \phi} )+ \frac{\Delta_0^2}{E_{k+p}+E_{k}}(\frac{\partial \varepsilon_{k}}{\partial \phi}  \frac{\epsilon_{k+p}}{E_{k+p} E_{k}^3}) \nonumber \\
\tilde{\Gamma}_{\phi \psi^* \psi}(p) &=&2 C_1 E_F \int \frac{d \theta}{2 \pi} \int d{\varepsilon}  \frac{-1}{(E_{k+p}+E_{k})^2}\frac{\Delta_0^2}{E_{k+p}E_{k}}( \frac{\varepsilon_{k}}{E_{k}} \frac{\partial \varepsilon_{k}}{\partial \phi} )+ \frac{-\Delta_0^2}{E_{k+p}+E_{k}}(\frac{\partial \varepsilon_{k}}{\partial \phi}  \frac{\epsilon_{k}}{E_{k+p} E_{k}^3}).
\end{eqnarray}
The energy integration requires a cut-off, $\Lambda$, and we assume it is much bigger than SC pairing. For the nematic coupling, we write
\begin{eqnarray}
\frac{\partial \varepsilon_k}{\partial \phi} =g \gamma(k^2) \cos(2\theta_k),
\end{eqnarray}
where our third parameter appears in  $\gamma(k^2)$
\begin{eqnarray}
\gamma(k^2) = 1+ C_3\frac{\epsilon_k}{E_F}.
\end{eqnarray}
For fermions with a circular Fermi surface ($\varepsilon(k) = k^2/(2m)-E_F$),  the introduced constants are
\begin{eqnarray}
C_1 =\frac{1}{\pi}  \quad C_2=\frac{1}{4} \quad C_3= 1.
\end{eqnarray}
At the BCS limit, the gap function becomes negligible, but it is dangerous to set it to be zero in the integration because quasiparticles' pole structures are affected by the gap magnitude. 

\begin{figure}
\includegraphics[width=5.0in]{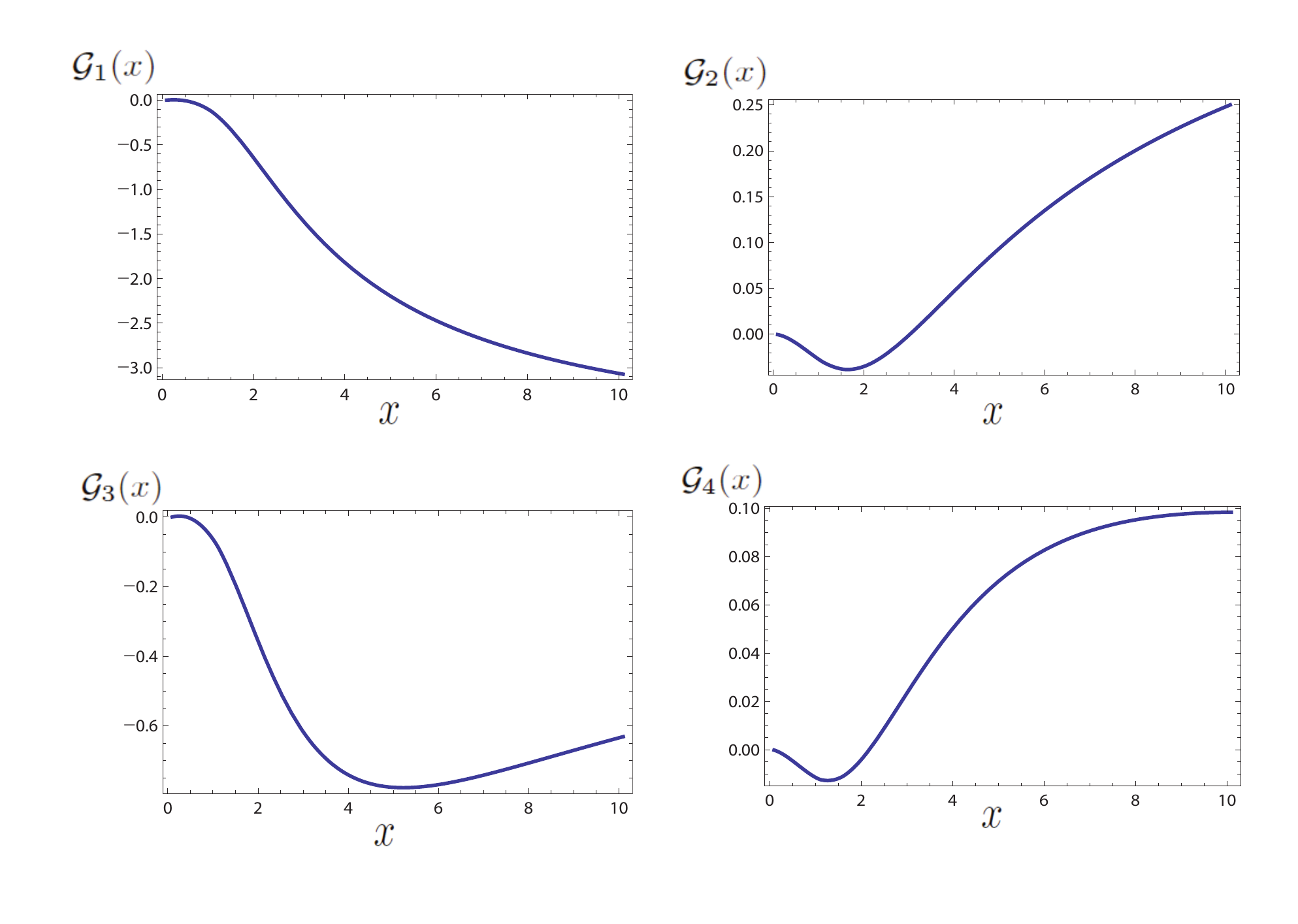}
\caption{Dimensionless functions of the vertex functions. See Eqns. \ref{ftns}.  } \label{functions}
\end{figure}

Note that we do not place any constraints on $p/\Delta$. 
By evaluating the vertex functions at the first oder in $\Delta/E_F$, we find that
\begin{eqnarray}
\Gamma_{\phi \psi^* \psi} (p)&=& 2 C_1 \frac{p_x^2-p_y^2}{p^2}(C_2 \mathcal{G}_1(p/\Delta) +  C_3 \mathcal{G}_2 (p/\Delta) ) \nonumber \\
\tilde{\Gamma}_{\phi \psi^* \psi} (p)&=& 2 C_1 \frac{p_x^2-p_y^2}{p^2}(C_2 \mathcal{G}_3(p/\Delta) + C_3 \mathcal{G}_4 (p/\Delta) ). \label{ftns}
\end{eqnarray}
It is not easy to get analytical forms for the vertex functions, so we illustrate the $\mathcal{G}_i$ functions in Fig.~\ref{functions}.
By performing asymptotic expansions, we obtain two limiting cases. 
The first limit is a metallic limit, $p/\Delta \rightarrow \infty$.
\begin{eqnarray}
&&\Gamma_{\phi \psi^* \psi} (p) \rightarrow const \nonumber \\
&&\tilde{\Gamma}_{\phi \psi^* \psi} (p) \rightarrow 0 
\end{eqnarray}
The other limit is a deep SC limit, $p/ \Delta \rightarrow 0$.
\begin{eqnarray}
&& \Gamma_{\phi \psi^* \psi} (p) \sim p^2/\Delta^2 \nonumber \\
&& \tilde{\Gamma}_{\phi \psi^* \psi} (p) \sim p^2/\Delta^2.
\end{eqnarray}

Several remarks are in order. 
First of all, the Fermi energy scale from density of states in the vertex functions dropped out as expected. This is because we are considering a non-zero angular momentum channel in the vertex function, so the zero-momentum channel cancelled out. 
Next, the introduced three constants depend on microscopic details significantly. 
For example, different band structures from different lattices(or hopping parameters) change Fermi velocity and density of states, but one can absorb such differences into redefinition of the constants.
Moreover, different order parameter structures such as different SC pairings(or different origin of nematicity) will affect the $\Gamma_{\phi \psi^* \psi}$ function evaluation.
However, it is expected that final results will have the same functional behaviors with the adjusted constants. 
In other words, the functional behavior does not depend on details of mechanisms; different pairings like d-wave or s-wave and origin of nematicity only changes the three constants. 

\begin{figure}
\includegraphics[width=5.0in]{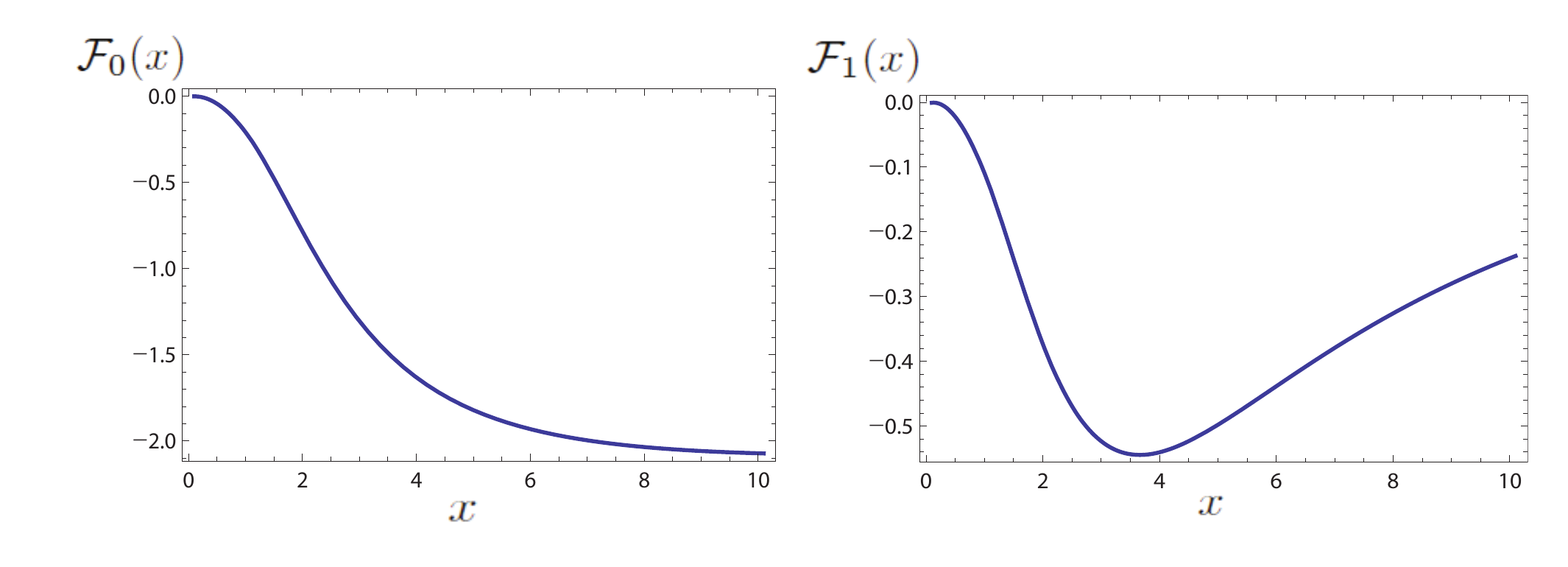}
\caption{total vertex functions for the spherical Fermi surface. $\mathcal{F}_0(p)$ and $\mathcal{F}_1(p)$ are defined in Eqn. \ref{spherical}} \label{tot}
\end{figure}

In Fig.~\ref{tot}, we illustrate the vertex functions for the spherical Fermi surface case. 
\begin{eqnarray}
&&\Gamma_{\phi \psi^* \psi} (p) = C_1 \mathcal{F}_0(p) \frac{p_x^2-p_y^2}{p^2} \nonumber \\
&&\tilde{\Gamma}_{\phi \psi^* \psi}(p)= C_1 \mathcal{F}_1(p) \frac{p_x^2-p_y^2}{p^2} \label{spherical}
\end{eqnarray}
As we can see, the functional behavior of the anisotropy vertex is almost monotonic.
The non-monotomic contribution comes from $g_3$ function which is much smaller than $g_1$. 
From that, we understand that decreasing SC indicates increasing anisotropy effect. 
It certainly shows competition between the nematic order and SC.  

Finally, let us estimate how much the non-analytic term makes the anisotropy enhanced inside the vortex core compared to the usual GL term contribution. 
For the sake of simplicity, let us assume SC is completely suppressed inside the vortex core, and has full gap size outside the vortex core. Of course, its size is determined by the coherence length (gap size).
Inside of the core, the functional behavior is approximated as 
\begin{eqnarray}
&&\Gamma_{inside} \sim \frac{1}{p^2} (p_x^2-p_y^2)
\end{eqnarray}
Then, the enhanced anisotropy inside the  vortex core, $\eta$, is basically 
\begin{eqnarray}
\eta = \int_{\xi^{-1}}^{a^{-1}} d^2 p  \frac{1}{p^2} \sim \log(\frac{\xi}{a})
\end{eqnarray}
up to some constant factors. We use a lattice length scale for comparison.  
This simple analysis shows that interplay between the Fermi surface and nematic order enhances the anisotropy of superconducting coherence lengths. 

\section{Discussion and Conclusion} \label{Conclusion}
Before closing the paper, let us take one step back and think about the meaning of our calculation. 
In theoretical physics, one of the most powerful concept is a low energy effective theory. 
It is very powerful because higher energy degrees of freedom are ignored and the theory become significantly simplified.
Following the spirit, in a $s$ wave SC, it is common to forget about fermion excitations in low energy limits because of the pairing gap in fermion spectrum.
So, in our system,  it seems reasonable to use a GL theory  to describe systems because fermions are gapped.
 
However, our calculation explicitly shows that such consideration is not enough and it is dangerous to use the naive GL theory. 
{How can we understand such inconsistency?}
The answer is hidden in Fermi surface physics especially when multiple order parameters are considered. 
Even though the pairing makes fermion spectrum gapped, the SC order and the nematic order are coupled to each other with a finite momentum transfer.  
Once the finite momentum is bigger than the gap magnitude, fermions start to see Fermi surface physics and the energy gap becomes irrelevant. 
In terms of our calculation, it exactly corresponds to the metallic case, $ p\xi \gg 1$.

To see importance of Fermi surface physics further, let us consider SC pairing susceptibility with nematic order parameter in the metallic ground state.
\begin{eqnarray}
\chi^{\phi}(2p) = \int_k \frac{1}{\epsilon^{\phi}_{k+p}+\epsilon^{\phi}_{k-p}}(\theta(-\epsilon^{\phi}_{k+p})-\theta(\epsilon^{\phi}_{k-p})),
\end{eqnarray} 
where the superscript($\phi$) indicates presence of the nematic order parameter in band structures.
The denominator basically describes particle-particle channel energy difference and the step functions are for allowed phase space contribution. 
If we take $p \rightarrow 0$ limit, then the usual BCS logarithmic instability shows up, but the presence of the finite momentum regularizes the susceptibility. 
The anisotropy SC coherence length vertex is basically $ \Gamma_{\phi \psi \psi^*} = {\partial \chi^{\phi}}/{\partial \phi}|_{\phi=0} $, 
 and it is easy to check it indeed corresponds to the form of the metallic limit.   

Based on these considerations,  the logic that gapped fermion theory only gives a GL theory is not correct.
One should keep in mind that possibility of Fermi surface effects. especially in the case where a momentum (or frequency) dependent function is present. 
Such non GL type term is not universal, and depends on order parameter properties significantly. 

It is interesting to compare our results with recent numerical works on the Hubbard model. \cite{tremblay, su}
It was reported that even small anisotropic hopping term with strong interaction induces strong enhancement of anisotropy in some physical quantities such as conductance. 
The enhancement basically comes from interplay physics between SC, anisotropic hopping, and fermion excitations, and it seems the numerical calculations are consistent with our results.  
It would be very interesting if one can find a mapping between the current work and reported numerical results. 

So far, we mainly focus on a low energy theory to understand general competition effects between the two order parameters.
Let us briefly comment on microscopic mechanisms on origin of nematicity in $FeSe$. 
First, it is unlikely  for  structural orthorhombicity alone to be an origin of strong vortex shape nematicity because it is too small even with non-analytic contribution from Fermi surfaces.  
Thus, electronic nematicity is likely to happen. Possible origins for the electronic nematicicty were suggested in literatures.\cite{cenke, song, rafael}
Magnetic ordering could be a candidate, but it was reported  that there is no hole pocket and the nesting wave vectors ,$(\pi,0)$ or $(0,\pi)$, are supressed in $FeSe$.\cite{micro,dai}
Alternatively, orbital ordering does no require any nesting, so it was proposed as another candidate for the nematicity. However, it is still an open question and remains to be understood in future works. No matter what is the microscopic mechanism of the nematicity, our main point is that magnitude of the nematic order is not required to be very big to explain the strongly distorted vortex shape. 

In this paper, we study interplay physics between nematic order and superconductivity. 
We show that competition between the order parameters is well captured by incorporating Fermi surface excitations coupling to the order parameters.
It is shown that anisotropy of superconducting coherence lengths is descriebd by the non-local term induced by fermion excitations.
We show that the non-local term becomes more important in metallic limit, which shows $(p_x^2-p_y^2)/p^2$ behavior in the SC coherence anisotropy vertex function.
Therefore, we argue the anisotropy channel becomes more enhanced near the vortex core, where SC is significantly suppressed. 

\acknowledgements

We thank E.~Berg and D.~Chowdhury for useful discussions.
This research was supported by the National Science Foundation under grant DMR-1103860 and by a MURI grant from AFOSR.

\end{document}